\documentclass[]{article}

\def\C{\mathbb{C}} 
\def\G{\mathbb{G}} 
\def\Q{\mathbb{Q}}
\def\R{\mathbb{R}}  

\def\no{\noindent}

\def\beq{\begin{equation}}
\def\eeq{\end{equation}}

\usepackage{array}
\usepackage{tabularx}
\usepackage{amsfonts}
\usepackage{amssymb}
\usepackage{tikz-cd}
\usepackage{graphicx}   
\def\bpm{\begin{pmatrix}}
\def\epm{\end{pmatrix}}
\makeindex            
\begin{document}

\title{Scaffolding of Spacetime}
\author{Garret Sobczyk
\\ Universidad de las Am\'ericas-Puebla
 \\ Departamento de F\'isico-Matem\'aticas
\\72820 Puebla, Pue., M\'exico
\\ http://www.garretstar.com}
\maketitle
\begin{abstract}The mathematical foundations of relativistic quantum mechanics is largely based upon the discovery of the Pauli and Dirac matrices. An algebra which lies
	at an even more fundamental level is the geometric Clifford algebra with metric signature (2,3)=(++---). In this geometric algebra both the fundamental Pauli vectors of space and the Dirac vectors of spacetime are factored into complex bivectors. 
	
\smallskip
\no {\em AMS Subject Classification:} 15A63, 15A66, 81T10, 81Q99
\smallskip

\no {\em Keywords:} Clifford algebra, complex numbers, geometric algebra, spacetime algebra, Witt basis. 
\end{abstract}
\newtheorem{thm1}{Theorem}
\newtheorem{thm2}[thm1]{Theorem}
\section{Geometric matrices}

The {\it geometrization of the real number system} \cite{Sgeo17}, is captured in the following
\begin{quote}{\bf Axiom:}   The real number system can be geometrically extended to include new,
anti-commutative square roots of $\pm 1$, each new such root representing
the direction of a unit vector along orthogonal coordinate axes of a Euclidean or
pseudo-Euclidean space $ \R^{p,q}$, where $p$ and $q$ are the number of new square roots of $+1$ and $-1$, respectively. 
\end{quote}

\no The resulting real associative {\it geometric algebra}, denoted by
\[  \G_{p,q} := \R(e_1, \ldots, e_p, f_1, \ldots, f_q), \]
has dimension $2^{p+q}$ over the real numbers $\R$, and is said to be {\it universal} since no further relations between the new square roots are assumed \cite{SNF,S2019}.
Of particular interest here are the geometric algebras $\G_{2,2}:=\G(\R^{2,2})$ and
$\G_{2,3}=\G(\R^{2,3})$.  

By a {\it geometric matrix} over the real or complex numbers we mean a square matrix of dimension $2^n\times 2^n$ in the matrix algebra $M_F(2^n)$ where $F=\R$ or $\C$ \cite{S08}, although other fields can be considered \cite[pp.195-204]{LP97}. What is particularly important about these matrix
algebras is that they are the {\it coordinate matrices} of the elements in the {\it isomorphic} geometric algebras.       

To see how this works, we use a Witt basis of {\it null vectors},\footnote{By a null vector in a geometric algebra we mean a nilpotent vector of index 2.}
  \beq   \{a_1, \ldots, a_n,b_1,\ldots, b_n\}, \label{specwitt} \eeq
 where for $i\ne j$, $a_ia_j =-a_j a_i$, $b_ib_j=-b_j b_i$ and $a_i b_j=-b_j a_i$. In addition, for $1\le i\le n$, we assume the following two properties
\begin{itemize}
	\item[N1.] $a_i^2=0=b_i^2$, 
	
	\item[N2.] $a_i b_i + b_i a_i =1$ for $1\le i\le n$. 
\end{itemize}  
Amazingly, by these simple rules we have completely characterized matrix multiplication of the coordinate matrices of the geometric algebras now to be defined \cite{S2018}.

The simplest case is when $n=1$, the case of $2\times 2$ real and complex matrices and their geometric algebra counterparts $\G_{1,1}=\R(a,b)$ and $\G_{1,2}=\C(a,b)$, respectively, for $a=a_1$ and $b=b_1$.
The multiplication table for the Witt basis $\{a,b\}$ of these geometric algebras easily follows from these two rules and is given in Table \ref{table3.1}.
\begin{table}[h!]
	\begin{center}
		\caption{Multiplication table.}
		\label{table3.1}
		\begin{tabular}{c|c|c|c|c} 
			& $a $  & $ b $  & $a b $  & $ b a $ \cr 
			\hline
			$ a$  &  $ 0$   & $ a b $     & $  0 $         & $ a $ \cr 
			\hline         
			$ b $  &  $ b a $  & $ 0 $     & $  b $         & $ 0 $ \cr 
			\hline     
			$ a b $    &  $a $    &  0  &  $a b $    & 0  \cr
			\hline
			$ b a $  &  0   &  $b $   &  0     & $b a$   \cr
		\end{tabular}
	\end{center}
\end{table}  
For example, to calculate $(ab)(a)$, we multiply the equation in N2 on the left by $a$ to get
\[  a(ab+ba)=a^2b + aba =a,  \]
since by the N1 $a^2=0$.
    
By the {\it spectral basis} of null vectors (\ref{specwitt}) of the geometric algebra $\G_{1,1}$, we mean
   \beq 
   \pmatrix{1 \cr a}u\pmatrix{1 & b}=\pmatrix{ba \cr a}\pmatrix{ba &b}=\pmatrix{ba & b \cr a & ab},  \label{specg11} \eeq
 where $u = ba$, \cite{S1,S0}.   
A geometric number $g\in \G_{1,1}:=\R(a,b)$, with respect to this basis, has the form
\beq    g = g_{11}ba + g_{12}b +g_{21}a + g_{22}ab =\pmatrix{ba & a}[g]\pmatrix{ba \cr b} 
 \label{matrixg} \eeq
where $[g]:=\pmatrix{g_{11} & g_{12} \cr g_{21} & g_{22}}$, for $g_{ij}\in \R $, is the
real {\it coordinate matrix} of $g$, and $u=ba$.  The matrices of the
Witt basis of null vectors $\{a,b\}$ are
\[ [a]:=\pmatrix{0 & 0 \cr 1 & 0} \ \ {\rm and} \ \ [b]=\pmatrix{0 & 1 \cr 0&0},  \]    
and satisfy the same multiplication Table \ref{table3.1}, proving the algebra isomorphism  $M_2(\R)\widetilde =\G_{1,1}$.
From the spectral basis of null vectors (\ref{specwitt}), we construct the corresponding {\it standard orthonormal basis} $\{ e,f \}$ of the geometric algebra $\G_{1,1}=\R(e,f) $,
by defining $e:= a+b$ and $f:=a-b$. It is easily checked, using N1 and N2, with the help of Table \ref{table3.1}, that $e^2=1=-f^2$ and $ef=-fe$. 

In the case of the
geometric algebra 
\beq G_{1,2}=\R(e_1,f_1,f_2)\widetilde=\C(e_1,f_1) , \label{complexg11} \eeq
The {\it complex matrices} $[g]$, with $g_{ij}\in \C$, are the coordinate matrices of the geometric algebra $\G_{1,2}=\R(e_1,f_1,f_2)  $. In (\ref{complexg11}), we have only asserted that $\G_{1,2}$ is isomorphic to
the complex geometric algebra $\C(e_1,e_2)$. If we define the third orthonormal vector
\[ f_2 :=e_1f_1 i \quad \iff \quad i=\sqrt{-1}=e_1f_1 f_2, \]
identifying $i$ as the pseudoscalar of $\G_{1,2}$, then we can make the stronger assertion that
\beq \G_{1,2}=\R(e_1,f_1,f_2)= \C(e_1,f_1). \label{strongg23} \eeq
 It is easily checked that $f_2^2=-1$, and $f_2$ anticommutes with
 $e_1$ and $f_1$. 
 
 The {\it Pauli algebra} ${\cal P}_3$, closely related to the geometric algebra $\G_3$ of Euclidean space $\R^3$, is defined by
 \[ {\cal P}_3=\R(\sigma_1,\sigma_2,\sigma_3),     \]
 where the Pauli vectors $\sigma_k$ are anti-commutative and satisfy
 \[  \sigma_1^2=\sigma_2^2=\sigma_3^2=1.\]
 Alternatively, the Pauli algebra ${\cal P}_3$ can be defined in terms of the geometric
 algebra $\G_{1,2}$. We have
 \beq {\cal P}_3(e_1) =\R(e_1,e_1f_2,e_1f_1)=\R(\sigma_1,\sigma_2,\sigma_3)  \widetilde=\G_{1,2}  \label{altpauli} \eeq 
for
 \[  \sigma_1:=e_1, \ \sigma_2:=e_1f_2, \ \sigma_3:=e_1f_1.   \]
  The coordinate matrices $[\sigma_k]$ of the Pauli vectors $\sigma_k$ are the famous Pauli matrices,
 \beq [\sigma_1]=\pmatrix{0 & 1\cr 1 & 0},\ [\sigma_2]=\pmatrix{0 & -i \cr i  & 0}, \ [\sigma_3]=\pmatrix{1 & 0 \cr 0 & -1}.   \label{fampauli} \eeq  
 
 By the spectral basis of real and complex, null vectors (\ref{specwitt}) for the geometric algebras
 $\G_{n,n}$ and $\G_{n,n+1}$, respectively, we mean
 \beq  A_1\overrightarrow{\otimes} \cdots \overrightarrow{\otimes}A_n u_{1\cdots n} B_n^T \overleftarrow{\otimes} \cdots \overleftarrow{\otimes}B_1^T,  \label{eqn1} \eeq 
 where we are employing the {\it directed left} and {\it right Kronecker products} $\overleftarrow{\otimes}$, $\overrightarrow{\otimes}$, respectively, and
 \[ A_i:=\pmatrix{1 \cr a_i} \ \ {\rm and} \ \ B_i^T=\pmatrix{1 & b_i},   \]
for $1 \le i \le n$ and $u_{1\cdots n}=u_1\cdots u_n$ for $u_i:=b_i a_i$. Just as for the case when $n=1$, which we have already given in (\ref{specg11}) and (\ref{matrixg}), the real or complex coordinate {\it geometric matrices} $[g]$ for a geometric number $g\in \G_{n,n}$ or $g\in \G_{n,n+1}$, respectively, are defined by the equation
\beq g=  A_1^T\overrightarrow{\otimes} \cdots \overrightarrow{\otimes}A_n^T u_{1\cdots n}[g] B_n \overleftarrow{\otimes} \cdots \overleftarrow{\otimes}B_1 .  \label{eqn2} \eeq

The above equations (\ref{eqn1}) amd (\ref{eqn2}) establishes the isomorphisms between the geometric coordinate matrices and the geometric algebras,
\beq \G_{n,n}=\R(e_1,\cdots,e_n,f_1,\cdots, f_n ) \widetilde =M_{2^n}(\R) \label{stdesfs} \eeq
and
\beq \G_{n,n+1}\widetilde=\C(e_1,\cdots,e_n,f_1,\cdots, f_n )\widetilde = \G_{2n}(\C)  \widetilde =M_{2^n}(\C),    \label{stdcomplexef} \eeq
where $e_i:=a_i+b_i$ and $f_i:=a_i-b_i$. In the complex case, when the pseudoscalar
$j=e_1f_1 \cdots e_nf_n f_{n+1}$ of $\G_{n,n+1}$, in the center of the algebra, is set {\it equal} to $\sqrt{-1}$, and $f_{n+1}:=\sqrt{-1} e_1f_1\cdots e_nf_n$, the stronger relationship 
\beq    \G_{n,n+1}=\R(e_1,\ldots, e_n,f_1, \cdots, f_{n+1})=\C(e_1,\cdots,e_n,f_1,\cdots,f_n)   \label{gnnplusone} \eeq
is valid \cite{S2019}. However, it must be remembered when using this relationship 
that $\sqrt{-1}=e_1f_1e_2f_2f_3$ plays the role of the imaginary number $\sqrt{-1}\in \C$.  

The case for $n=2$ in equations (\ref{eqn1}) and (\ref{eqn2}) is treated in the next section.

\section{Scaffolding of spacetime} 
In this section we give the details of the construction of the geometric algebra $\G_{2.3}$, providing a scaffolding for both the Dirac and Pauli algebras.  
The basic equations (\ref{eqn1}) and (\ref{eqn2}), for $n=2$, give the geometric algebra
\[ \G_{2,3}=\R(e_1,e_2,f_1,f_2,f_3)=\C(e_1,e_2,f_1,f_2)\widetilde{=} M_{2^2}(\C) , \]
where we have identified $f_3=je_1f_1e_2f_2$ for $j\equiv e_1f_1e_2f_2f_3\equiv\sqrt{-1}$.

The equation (\ref{eqn1}) for the spectral basis of complex null vectors (\ref{specwitt}) of $\G_{2,3}$, gives
\[  A_1\overrightarrow{\otimes}A_2 u_{12} B_2^T \overleftarrow{\otimes}B_1^T =\pmatrix{1 \cr a_1 \cr a_2 \cr a_{12}}
u_{12}  \pmatrix{1 & b_1 & b_2 & b_{21}} \] 
\beq= \pmatrix{u_{12} & b_{1} u_2 & b_{2}u_1 & b_{21} 
	\cr a_1 u_2 & u_1^\dagger u_2 & a_1 b_2 &-b_2 u_1^\dagger   \cr
	a_{2}u_1  & a_2 b_1 & u_1 u_2^\dagger & b_1 u_2^\dagger  \cr
	a_{12} & -a_2 u_1^\dagger & a_1u_2^\dagger & u_{12}^\dagger}, \label{eqn1g23} \eeq 
where $u_i:= b_ia_i$, $u_i^\dagger:=a_i b_i$, $u_{12}:=u_1u_2$, and $u_{12}^\dagger := u_2^\dagger u_1^\dagger=u_1^\dagger u_2^\dagger$. For a complex coordinate matrix $[g]$, 
the corresponding geometric number $g\in \G_{2,3}$, is given by the second equation (\ref{eqn2}),  
\beq g=  A_1^T\overrightarrow{\otimes} A_2^T u_{12}[g] B_2 \overleftarrow{\otimes}B_1 =\pmatrix{1 & a_1 & a_2 & a_{12}}
u_{12}[g]  \pmatrix{1 \cr b_1 \cr b_2 \cr b_{21}}, \label{eqn2g23} \eeq
where $[g]$ is the complex coordinate $4\times 4$ matrix of $g\in \G_{2,3}$. 

We now give pertinent relationships between geometric algebras and their even subalgebras, which are used to model the structure of spacetime. The {\it Dirac Algebra} $\G_{1,3}$ of gamma vectors $\{\gamma_\mu\}$, for $\mu = 0,1,2,3$, are defined by
\beq \G_{1,3}=\R(\gamma_0,\gamma_1,\gamma_2,\gamma_3) \label{DGA} \eeq
where 
\[ \gamma_0^2=1, \quad {\rm and} \quad \gamma_1^2=\gamma_2^2=\gamma_3^2=-1 \]
and are pare-wise anticommutative. Hestenes' {\it spacetime split} \cite[p.25]{H66} of the Dirac algebra into the Pauli algebra ${\cal P}_3(\gamma_0)$, by the timelike vector $\gamma_0$, is simply the recognition 
that 
\beq \G_{1,3}^+\widetilde=\R(\gamma_{10},\gamma_{20},\gamma_{30})=\R(\sigma_1,\sigma_2,\sigma_3) ={\cal P}_3(\gamma_0),     \label{g13plus1}  \eeq
where $\sigma_k := \gamma_k \gamma_0$ for $k=1,2,3$.
It follows from (\ref{altpauli}) and (\ref{g13plus1}) that 
\beq {\cal P}_3(e_1)\widetilde =\G_{1,2}\widetilde ={\cal P}_3(\gamma_0).     \label{g13plus2} \eeq
The first relationship shows that the Pauli algebra is isomorphic to $\G_{1,2}$, whereas the second relationship shows that the even subalgebra of the Dirac algebra $\G_{1,3}$ is also isomorphic to $\G_{1,2}$.

Regarding the even subalgebras of the geometric algebra $\G_{2,3}$, we have for $j=e_1f_1e_2f_2f_3=\sqrt{-1}$,
\beq \G_{1,3}(f_1):=\G(\R^{1,3}) \widetilde = \G_{2,2}^+(\C)=\R(e_{1}f_1,je_2f_1,f_{12},f_{31})\widetilde= M_2(\Q), \label{g23plus13}, \label{g23plus1} \eeq     
\beq \G_{3,1}(e_1):=\G(\R^{3,1}) \widetilde=  \G_{2,2}^+(\C)=\R(je_{21},f_1e_1,f_2e_1,f_3e_1)\widetilde= M_4(\R). \label{g23plus31} \eeq  
\beq \G_4(e_1):=\G(\R^{4}) \widetilde=  \G_{2,2}^+(\C)=\R(e_{1}f_1,je_{21},f_2e_1,f_3e_1)\widetilde= M_4(\R). \label{g23plus4} \eeq 
\beq \G_{0,4}(f_1):=\G(\R^{0,4}) \widetilde=  \G_{2,2}^+(\C)=\R(je_{1}f_1,je_2f_1,f_{12},f_{31})\widetilde= M_2(\Q), \label{g23neg4} \eeq 
where $\Q$ denotes the {\it quaternions}. Each of the above pseudo-Euclidean spaces $\R^{p,q}$ is modeled by an even subalgebra of complex bivectors of the geometric algebra $\G_{2,2}^+(\C)$. See (\ref{g23plus13}) through (\ref{g23neg4}), respectively. We will only discuss (\ref{g23plus13}) in the next Section.

\section{Factoring the Dirac and Pauli algebras}

If we consider geometric algebras over the complex numbers, the classification of the algebras becomes much simpler, \cite[p.217]{LP97}, \cite{S2019}. In (\ref{strongg23}), we saw that the algebra $\G_{2,3}$ is already complexified when its pseudoscalar, the unit $5$-vector
$j=e_1f_1e_2f_2f_3 =\sqrt{-1}$, so that $\G_{2,3}=\G_{2,2}(\C)$. 

Let us study now in detail the relationship (\ref{g23plus1}),
\beq \G_{1,3}(f_1)=\R(\gamma_0,\gamma_1,\gamma_2,\gamma_3)=\R(e_{1}f_1,je_2f_1,f_{12},f_{31})=  \G_{2,2}^+(\C), \label{splittinging23} \eeq
where $\gamma_0:=e_1f_1,\gamma_1:=je_2 f_1,\gamma_2:=f_{12},\gamma_3:=f_{31}$. For the Pauli vectors, we then find that 
\beq \sigma_1=\gamma_{10}=je_{21},\ \sigma_2=\gamma_{21}=e_1f_2,\ \sigma_3=\gamma_{31}=f_{3}e_1 \label{doublesplit} \eeq   
We saw in (\ref{g13plus1}) how the Dirac algebra is split into a corresponding algebra of Pauli vectors (Dirac bivectors), determined uniquely by the timelike Dirac vector $\gamma_0$, \cite{H66}. Similarly, (\ref{splittinging23}) and
(\ref{doublesplit})
represents a {\it double splitting} of the geometric algebra $\G_{2,3}=\G_{2,2}(\C)$, by the vectors $f_1=\gamma_{0123}$ and $e_1=\gamma_{321}$, of both the Dirac vectors $\{\gamma_\mu\}$ by $f_1$ and the Pauli vectors $\{\sigma_k\}$ by $e_1$. The hope of such a double splitting in the larger space $\G_{2,3}$ is that it leads to new insights into the foundations of relativistic quantum mechanics, \cite{AH02}.   

By {\it redefining} the matrices of the null vectors $a_i$ and $b_i$ to be
\[  [a_1]:=\pmatrix{0 & 0 & 0 & 0 \cr 0&0&0& 0 \cr i & 0&0&0 \cr 0& i & 0 &0} ,\  [b_1]:=[a_1]^*=\pmatrix{0 & 0 & -i & 0 \cr 0&0&0&-i \cr 0 & 0&0&0 \cr 0& 0 & 0 &0}  \] 
\[  [a_2]:=\pmatrix{0 & 0 & 0 & 0 \cr 1&0&0& 0 \cr 0 & 0&0&0 \cr 0& 0 & -1 &0} ,\  [b_2]:=[a_2]^*=\pmatrix{0 & 1 & 0 & 0 \cr 0&0&0&0 \cr 0 & 0&0&-1 \cr 0& 0 & 0 &0},  \]
the matrices of the Dirac vectors $\gamma_\mu$ and Pauli vectors $\sigma_k$, given in
(\ref{splittinging23}) and (\ref{doublesplit}) in terms of the $e_i,f_i$ defined in (\ref{stdcomplexef}), become the standard Dirac and Pauli matrices used by physicists \cite[p.110]{Green1968}, \cite{wikigamma}.
We have
\beq [\gamma_{10}]=\pmatrix{[1]_2 & [0]_2 \cr [0]_2 & -[1]_2}, \ \ {\rm and} \ \ [\gamma_{k0}]=\pmatrix{[0]_2 & -[\sigma_k]_2 \cr [\sigma_k]_2&[0]_2},   \label{tDirac} \eeq
where $[\sigma_k]_2$ are the Pauli matrices (\ref{fampauli}) for $k=1,2,3$. Also, 
\[ \gamma^5:=i f^1=-i f_1=e_{12}f_{32} = \pmatrix{0&0&1&0\cr 0&0&0&1 \cr 1&0&0&0\cr 0&1&0&0} \] 
is traditionally identified as the Dirac $\gamma^5$ vector \cite{wikigamma}, so the $\gamma^5$ is the 
{\it dual} of $f_1=\gamma_{0123}$ with respect to the $5$-vector $i=e_1f_1e_2f_2f_3$. 

However, there is something going on in (\ref{fampauli}) and (\ref{tDirac}) that is at worst, confusing. The $i=\sqrt{-1}$ in the Dirac matrices is the $5$-vector $e_1f_1e_2f_2f_3\in \G_{2,3}$, whereas the $i=\sqrt{-1}$ in the Pauli matrices (\ref{fampauli}) is the $3$-vector $\sigma_{123}=e_1f_{12}\in \G_{1,2}$. 
At best, the double-splitting of $\G_{2,3}$ into the Dirac and Pauli algebras will lead to new insights.

\section*{Acknowledgements} I thank Professor Melina Gomez Bork for stimulating discussions related to this work, and the Universidad de Las Americas-Puebla for many years of support.

\end{document}